\documentclass[prd,aps,nofootinbib,floatfix,11pt]{revtex4}
\usepackage{amsmath,graphicx,epsfig,amssymb,dsfont,mathtools,}
\usepackage[usenames]{color}
\usepackage{ulem} %% for strike-through
\usepackage{bigstrut}
\usepackage{slashed}
\usepackage{multirow}
\usepackage{subfigure}

\allowdisplaybreaks

\begin{document}
\title{Factorization formula connecting the $\Lambda_Q$ LCDA in QCD and boosted HQET} %within the method-of-regions}

\author{Yu-Ji Shi$^{1}$~\footnote{Email: shiyuji@ecust.edu.cn}, Jun Zeng~$^{2,3}$~\footnote{Email: zengj@hainnu.edu.cn (Corresponding author)}}
\affiliation{$^1$ School of Physics, East China University of Science and Technology, Shanghai 200237, China\\
  $^{2}$ College of Physics and Electronic Engineering, Hainan Normal University, Haikou 571158, Hainan, China\\
  $^{3}$ State Key Laboratory of Dark Matter Physics, Key Laboratory for Particle Astrophysics and Cosmology (MOE), Shanghai Key Laboratory for Particle Physics and Cosmology, School of Physics and Astronomy, Shanghai Jiao Tong University, Shanghai 200240, China}

\date{\today}
\begin{abstract}
Light-cone distribution amplitudes (LCDAs) are essential to precision phenomenology in heavy baryon decays. In this work, we derive a factorization formula connecting the leading-twist QCD LCDA to the boosted HQET LCDA of the $\Lambda_Q$ baryon in the peak region. We demonstrate a significant simplification of the matching procedure by applying the method-of-regions to perturbative calculations. With this simplification, we calculate the required one-loop perturbative corrections to the QCD and boosted HQET LCDAs in the $\overline{\rm MS}$ scheme, and thereby obtain the one-loop jet function that serves as the matching kernel in the factorization formula. This result provides a critical step toward lattice QCD calculation of heavy baryon LCDAs in the future.
\end{abstract}
\maketitle

\section{Introduction}
In recent years, weak decays of heavy baryons have emerged as a frontier area in flavor physics, driven by the search for direct $CP$ violation (CPV) in heavy baryon decays. This direction was greatly accelerated by a significant breakthrough in 2025, when the LHCb collaboration reported evidence of CPV in the decay $\Lambda_b^0 \to pK^-\pi^+\pi^-$~\cite{LHCb:2025ray}, a channel previously highlighted by a theoretical prediction in Ref.~\cite{Han:2024kgz}.
Theoretically, heavy baryon decays present greater challenges compared to heavy meson decays. The presence of three valence quarks renders the factorization picture much more intricate than the meson case. 
Nowadays, theoretical predictions for $\Lambda_b$ decays rely on factorization-based approaches such as QCD factorization (QCDF)~\cite{Beneke:1999br,Beneke:2000ry}, soft-collinear effective theory (SCET)~\cite{Bauer:2000yr,Bauer:2001yt,Bauer:2002nz,Wang:2011uv,Lu:2025gjt}, perturbative QCD (PQCD)~\cite{Shih:1998pb,Keum:2000wi,Lu:2000em,Keum:2000ph,Lu:2009cm,Han:2022srw,Han:2025tvc,Li:2025rsm,Yang:2025yaw} and light-cone sum rules (LCSR)~\cite{Wang:2009hra,Wang:2015ndk,Miao:2022bga,Khodjamirian:2023wol,Huang:2024oik}. These methods systematically separate the perturbative hard scattering from the non-perturbative dynamics, where the $\Lambda_b$ light-cone distribution amplitude (LCDA) serves as the essential non-perturbative input. The LCDA encodes the partonic structure of the baryon and directly enters the calculation of decay amplitudes and CP-violating observables. Therefore, a precise determination of the $\Lambda_b$ LCDA is crucial for reliable theoretical predictions in heavy baryon decays.

A first-principles determination of heavy hadron LCDAs faces two main theoretical difficulties. First, as LCDAs are defined on the light-cone, their direct computation using lattice QCD is inherently difficult. Second, heavy baryon LCDAs are formulated within heavy‑quark effective theory (HQET) and involve the effective heavy‑quark field $h_v$, introducing additional technical complications for lattice simulations. 
In recent years, a two-step matching scheme  has successfully addressed both of these challenges in the context of heavy mesons LCDAs~\cite{Han:2024fkr,LatticeParton:2024zko}, and the same strategy can also be applied to heavy baryons. This framework employs two sequential effective field theories. The first step relies on Large Momentum Effective Theory (LaMET)~\cite{Ji:2013dva,Ji:2014gla,Ji:2020ect,Cichy:2018mum}. Instead of light‑cone quantities, one defines an equal‑time correlation function (quasi DAs) of the heavy baryon carrying a large momentum $P^z$.  After factorizing the hard and collinear modes, one arrives at the QCD LCDAs encoding the dynamics at scales much lower than $P^z$.  In the second step, collinear modes with off‑shellness of heavy quark mass square $m_Q^2$ are integrated out, and the QCD LCDAs are matched onto the HQET LCDAs in a boosted frame.

The quasi‑DAs are matched to the QCD LCDAs through a factorization formula, in which the quasi‑DAs are written as a convolution of a short‑distance matching kernel with the QCD LCDAs. At leading power, the matching kernel is independent of the heavy-quark mass. Consequently, the leading-power matching kernel between quasi-DAs and QCD LCDAs for a heavy baryon is identical to that for a light baryon of the same spin, a result that has been derived in Ref.~\cite{Deng:2023csv, Han:2024ucv}.
Therefore, the key step is to match the QCD LCDAs to the HQET LCDAs in a boosted frame, where the the QCD LCDAs is described by soft-collinear effective theory (SCET), and the HQET LCDAs is described by boosted HQET (bHQET) \cite{Dai:2021mxb}. It is clear that the QCD LCDAs receives contributions from both the short-distance scale of $m_Q$ and the long-distance scale of $\Lambda_{\text{QCD}}$. The separation of these two scales gives rise to large logarithms of the form ${\rm ln} (m_Q / \Lambda_{\text{QCD}})$, which must be properly resummed using renormalization group (RG) techniques. This resummation is achieved by factorizing the QCD LCDAs into a convolution of a perturbative jet function at the scale $m_Q$ with the bHQET LCDAs~\cite{Ishaq:2019dst,Beneke:2023nmj}. In this framework, the momentum fractions of the light quarks in QCD LCDAs are divided into two distinct regions: a small-fraction region corresponding to the peak of the LCDAs, and a large-fraction region associated with its tail. Since the momentum fractions of the light quarks are typically much smaller than that of the heavy quark, the QCD LCDAs in the tail region is not related to the bHQET LCDAs . Thus we will focus on the factorization formula connecting the QCD and bHQET LCDAs in the peak region.

In Ref.~\cite{Beneke:2023nmj}, the factorization in the peak region was performed by first calculating the complete one-loop corrections to the QCD and bHQET LCDAs, and then taking the small-momentum-fraction limit of the QCD result and matching it to the bHQET expression. Although this represents the standard matching procedure, a significant simplification can be achieved using the method-of-regions. Specifically, within the peak region, it can be demonstrated that most of the one-loop corrections to the QCD LCDAs do not contribute to the jet function. The feasibility of this method has been established in Refs.~\cite{Beneke:2023nmj,Deng:2024dkd}, but this method have not been applied to formally reproduce the jet function.

In this work, using the method-of-regions, we will derive a factorization formula connecting the leading twist QCD and bHQET LCDA for $\Lambda_{Q}$ baryon in the peak region. The remainder of this paper is arranged as follows. Sec.~\ref{sec:LCDAdefinition} presents the definition of the $\Lambda_{Q}$ LCDA in SCET and bHQET. Sec.~\ref{sec:methodofregions} provides an introduction to the factorization procedure using the method of regions. Sec.~\ref{sec:SCETcalculation} and Sec.~\ref{sec:bHQETcalculation} present the calculation of SCET and bHQET amplitudes. Sec.~\ref{sec:jetfunction} gives the result for the jet function. Sec.~\ref{sec:conclusion} is a conclusion of this work.

\section{Leading twist LCDA of $\Lambda_Q$ in SCET and bHQET}\label{sec:LCDAdefinition}

In the boosted frame, the leading-twist LCDA of the $\Lambda_Q$ baryon is defined within massive SCET, in close analogy to the LCDA of a boosted $\Lambda$ baryon presented in Ref.~\cite{Deng:2023csv}. The collinear part of the massive SCET Lagrangian is
\begin{align}
{\cal L_{{\rm SCET}}}=\bar{\xi}(x)\left[2i\bar{n}\cdot D+(i\slashed D_{\perp}-m_{Q})\frac{1}{in\cdot D}(i\slashed D_{\perp}+m_{Q})\right]\frac{\slashed n}{2}\xi(x),\label{eq:SCETLagra}
\end{align}
where $\xi=(\bar{\slashed n}\slashed n/2)Q$ is the large component of the heavy quark field.  The light-cone vector $n$ is defined as
\begin{align}
n^{\mu} & =\frac{1}{\sqrt{2}}(1,0,0,-1),\ \ \ \ \bar{n}^{\mu}=\frac{1}{\sqrt{2}}(1,0,0,1),\ \ \ \ n\cdot\bar{n}=1.
\end{align}
Here we use  the convention $p=(n\cdot p, p_\perp, \bar n \cdot p)=(p^+, p_\perp, p^-)$ to denote a four-vector in the light-cone coordinate.
In the boosted frame, the leading-twist QCD LCDA of $\Lambda_Q$ is expressed by the matrix element
\begin{align}
\Phi_{\rm QCD}(x_{1},x_{2})f_{\Lambda_{Q}}u_{\Lambda_{Q}}(p)=\langle0|{\cal O}_c(x_{1},x_{2})|\Lambda_{Q}(p)\rangle,\label{eq:defQCDLCDA}
\end{align}
where ${\cal O}_c(x_{1},x_{2})$ is a nonlocal operator defined as
\begin{align}
{\cal O}_c(x_{1},x_{2})= & (p^{+})^{2}\int\frac{dt_{1}}{2\pi}\int\frac{dt_{2}}{2\pi}e^{ix_{1}p^{+}t_{1}+ix_{2}p^{+}t_{2}}\nonumber\\
 & \times\epsilon_{ijk}W_{ii^{\prime}}(t_{1}n)u_{i^{\prime}}^{T}(t_{1}n)\Gamma W_{jj^{\prime}}(t_{2}n)d_{j^{\prime}}(t_{2}n)W_{kk^{\prime}}(0)Q_{k^{\prime}}(0).\label{eq:SCEToper}
\end{align}
The gauge link is set along the light-cone and takes the form
\begin{align}
W(t n)={\cal P}{\rm exp}\left[ig_s\int_{t}^{\infty}d\lambda\  n\cdot A^a(\lambda n) T^a\right].
\end{align}
 The leading twist component of the LCDA is projected out by the structure $\Gamma=C\gamma_5\slashed n$. $x_1, x_2$ are the momentum fractions of the $u, d$ quarks inside the $\Lambda_Q$ baryon. The decay constant $f_{\Lambda_{Q}}$ serves as the normalization factor for $\Lambda_Q$ LCDA. It is defined by the matrix element of a local operator:
 \begin{align}
f_{\Lambda_{Q}}u_{\Lambda_{Q}}(p)=\epsilon_{ijk}\langle0|u_i^T(0)\Gamma d_j(0)Q_k(0)|\Lambda_{Q}(p)\rangle. \label{eq:SCETlocalMatrix}
 \end{align}
 The momentum of $\Lambda_Q$ lies in the light-cone plane: $p^{\mu}=m_H v = p^{+} \bar{n}^{\mu} + (m_H^{2}/2p^{+}) n^{\mu}$, where $m_H$ is the baryon mass and $p^{+} \sim Q \gg m_H$ is a large energy scale.
 
 On the other hand, in the boost frame the dynamics of heavy quarks can be described by bHQET. In this frame, the heavy-quark momentum is parametrised as $p_Q = m_Q v + k$, where $v^{\mu} \sim (1/b, 1, b)$ is the velocity of the heavy baryon with the boost parameter $b = m_H/Q \ll 1$, and $k^{\mu} \sim (b, 1, 1/b) \Lambda_{\rm QCD}$ denotes the soft-collinear residual momentum. The Lagrangian of bHQET can be derived from that of massive SCET by the redefinition of the collinear field $\xi$ in Eq.~(\ref{eq:SCETLagra})
\begin{align}
\xi(x)=\sqrt{\frac{n\cdot v}{\sqrt{2}}}e^{-im_{Q}v\cdot x}h_{n}(x),\label{eq:bHQETfield}
\end{align}
where $h_{n}$ is the effective heavy quark field in the boosted frame, which is dominated by the soft‑collinear momentum $k$. The normalization factor is chosen so that the bHQET Lagrangian has the same form as that in Ref.~\cite{Dai:2021mxb}. Inserting Eq.~(\ref{eq:bHQETfield}) into Eq.~(\ref{eq:SCETLagra}), one can obtain the bHQET Lagrangian:
\begin{align}
    {\cal L}_{\rm bHQET}=\bar{h}_{n}(iv\cdot D)\frac{\slashed n}{\sqrt{2}}h_{n}+{\cal O}\left(\Lambda_{\rm QCD}/m_Q\right).\label{eq:bHQETLagran}
\end{align}

The leading-twist LCDA of $\Lambda_Q$ within bHQET is expressed by the matrix element
\begin{align}
\Phi_{\rm bHQET}(\omega_{1},\omega_{2}){\bar f}_{\Lambda_{Q}}u_{\Lambda_{Q}}(p)=\langle0|{\cal O}_h(\omega_{1},\omega_{2})|\Lambda_{Q}(p)\rangle,\label{eq:defHQETLCDA}
\end{align}
where ${\cal O}_h(\omega_{1},\omega_{2})$ is a nonlocal operator defined as \cite{Ball:2008fw,Bell:2013tfa}
\begin{align}
{\cal O}_h(\omega_{1},\omega_{2})= & (v^{+})^{2}\int\frac{dt_{1}}{2\pi}\int\frac{dt_{2}}{2\pi}e^{i\omega_{1}v^+ t_{1}+i\omega_{2}v^+ t_{2}}\nonumber\\
 & \times\epsilon_{ijk}W_{ii^{\prime}}(t_{1}n)u_{i^{\prime}}^{T}(t_{1}n)\Gamma W_{jj^{\prime}}(t_{2}n)d_{j^{\prime}}(t_{2}n)W_{kk^{\prime}}(0)h_{n, k^{\prime}}(0).
\end{align}
The projector $\Gamma$ and the gauge link $W(t n)$ are the same as that in Eq.~(\ref{eq:SCEToper}). The decay constant ${\bar f}_{\Lambda_{Q}}$ serves as the normalization factor defined by the matrix element of a local operator:
 \begin{align}
{\bar f}_{\Lambda_{Q}}u_{\Lambda_{Q}}(p)=\epsilon_{ijk}\langle0|u_i^T(0)\Gamma d_j(0)h_{n,k}(0)|\Lambda_{Q}(p)\rangle.\label{eq:bHQETlocalMatrix}
 \end{align}

Generally, the matching between ${\cal O}_c$ and ${\cal O}_h$ depends on the region of the light-cone momentum fractions $x_i$. The region $x_i \sim \Lambda_{\rm QCD}/m_Q$ corresponds to the parametric location of the peak of the LCDA and is therefore referred to as the peak region. Accordingly, the region where $x_i \sim 1$ is referred to as the tail region. The factorization formula matching ${\cal O}_c$ and ${\cal O}_h$ will take the form
\begin{align}
{\cal O}_{c}(x_{1},x_{2}) = \begin{cases}
{\cal J}_{\rm peak}(x_{1},x_{2},\omega_{1},\omega_{2}) \otimes {\cal O}_{h}(\omega_{1},\omega_{2}), & x_i \sim \Lambda_{\rm QCD}/m_Q, \\
{\cal J}_{\rm tail}(x_{1},x_{2}), &  x_i \sim 1,
\end{cases}\label{eq:OperMatching}
\end{align}
where $\otimes$ represents the convolution in $\omega_{1}$ and $\omega_{2}$. 
Since the momentum fractions $x_i$ of the light quarks are typically much smaller than that of the heavy quark, the QCD LCDA does not match the form of the bHQET LCDA in the tail region. Therefore, in this work we focus on the jet function in the peak region, denoted as $\mathcal{J}_{\rm peak}$.

\section{Matching with the method of regions}\label{sec:methodofregions}

\begin{figure}
\begin{center}
\includegraphics[width=0.9\columnwidth]{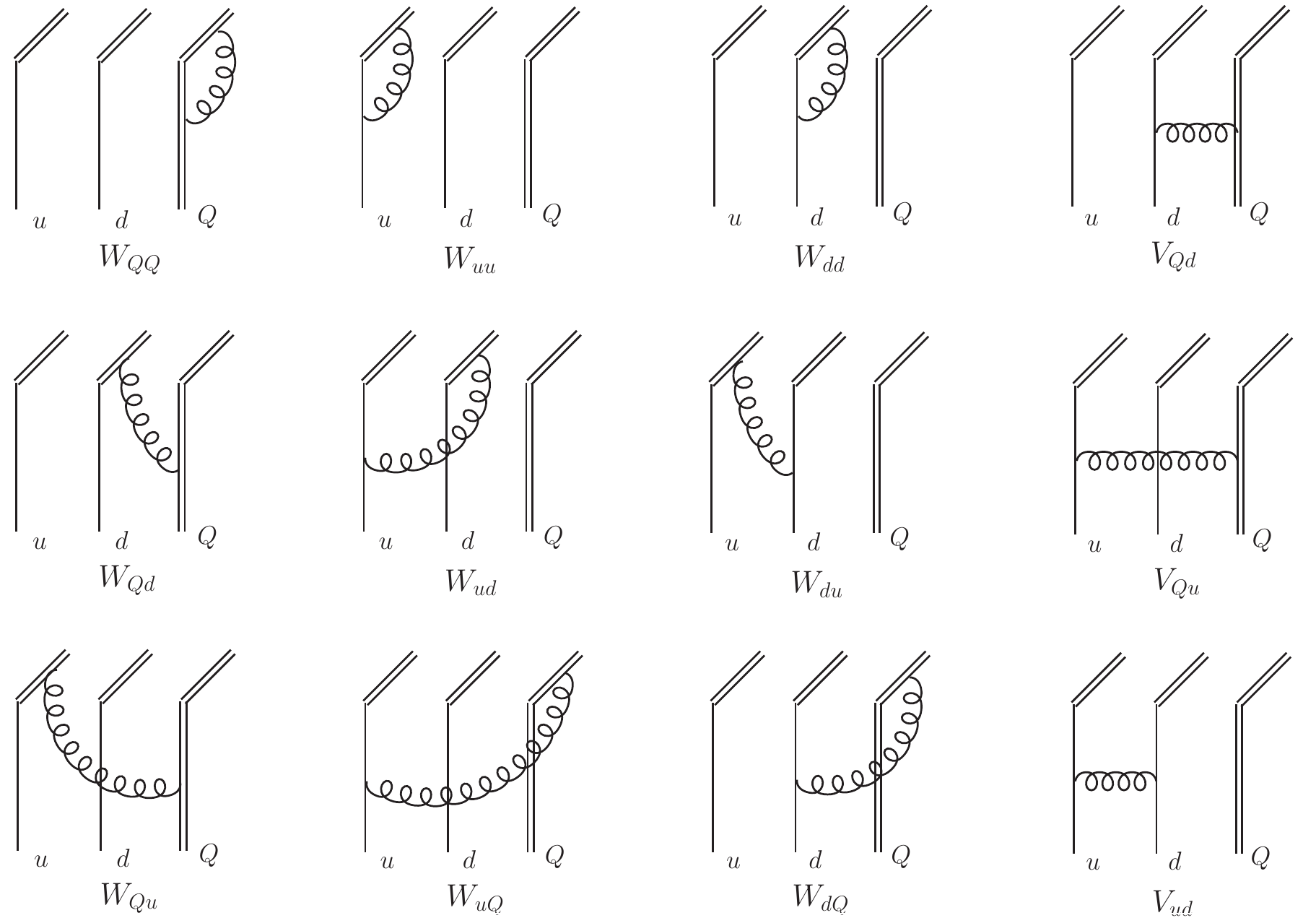} 
\caption{One-loop corrections to the matrix elements defined in Eq.~(\ref{eq:SCETmatrix1loop}) and Eq.~(\ref{eq:bHQETmatrix1loop}). The vertical single line denotes the light quark, the vertical double line represents the heavy quark, and the tilted double line corresponds to the gauge link attached to each quark fields.}
\label{fig:LCDAmatching} 
\end{center}
\end{figure}

Since the UV behavior of LCDAs is insensitive to low‑energy dynamics, one can match the operators ${\cal O}_c$ and ${\cal O}_h$ in the peak region by equating their matrix elements with external free‑quark states:
\begin{align}
&\langle0|{\cal O}_{c}(x_{1},x_{2})|{\overline {u(k_{1})d(k_{2})}}Q(p_{Q})\rangle\nonumber\\
=&\int d\omega_{1}d\omega_{2}{\cal J}_{\rm peak}(x_{1},x_{2};\omega_{1},\omega_{2})\langle0|{\cal O}_{h}(\omega_{1},\omega_{2})|{\overline {u(k_{1})d(k_{2})}}Q(p_{Q})\rangle,\label{eq:OcMatchOh} 
\end{align}
where the external momentums are set on-shell 
and we denote the momentums of $u,d,Q$ quarks as $k_1, k_2, p_Q$. $\overline{u(p_1)d(p_2)}$ denotes the spin‑averaged combination of the light quarks:
\begin{align}
    |{\overline {u(k_{1})d(k_{2})}}\rangle=\frac{1}{\sqrt{2}}\left(b_{u,\uparrow}^{\dagger}(k_{1})b_{d,\downarrow}^{\dagger}(k_{2})-b_{u,\downarrow}^{\dagger}(k_{1})b_{d,\uparrow}^{\dagger}(k_{2})\right)|0\rangle, \label{eq:spinaverage}
\end{align}
so that the $u$ and $d$ quarks form a total spin‑0 combination, as in the physical $\Lambda_Q$ baryon. On the SCET side, these momentums are parametrized as 
\begin{align}
p_{Q}=p_{Q}^{+}\bar{n}+\frac{m_{Q}^{2}}{2p_{Q}^{+}}n,\ \ \ \ k_{1,2}=k_{1,2}^{+}\bar{n},
\end{align}
with $k_{i}^{+}=x_{i,0}p^{+},\ \ \ p_{Q}^{+}=x_{3,0}p^{+}$. The corresponding matrix element up to ${\cal O}(\alpha_s)$ can be writen as
\begin{align}
&\langle0|{\cal O}_{c}(x_{1},x_{2})|{\overline {u(k_{1})d(k_{2})}}Q(p_{Q})\rangle\nonumber\\
    =&S^{(0)}\left[\delta(x_{1}-x_{1,0})\delta(x_{2}-x_{2,0})+\frac{\alpha_{s}C_{F}}{4\pi}M^{(1)}(x_{1},x_{2};x_{1,0},x_{2,0})\right],\label{eq:SCETmatrix1loop}
\end{align}
where $S^{(0)}$ is the tree level amplitude. $M^{(1)}$  contains the one-loop diagrams of SCET as shown in Fig.~\ref{fig:LCDAmatching}, ${\cal O}(\alpha_s)$ corrections to the light and residues of heavy quark field $Q$, as well as the renormalization factor canceling  UV divergence. Note that the matrix element at ${\cal O}(\alpha_s)$ is proportional to $S^{(0)}$. This follows from averaging over the light‑quark spins and will be discussed in the next section.

On the bHQET side, the quark momentums are parametrized as
\begin{align}
p_{Q} & =m_{Q}v+k_{Q},\ \ \ \ k_{1,2}=\nu_{1,2}v^+\bar{n},\nonumber\\
v^{\mu} & =v^+\bar{n}^{\mu}+\frac{1}{2v^+}n^{\mu},
\end{align}
where $k_Q$ is the soft-collinear momentum of the effective heavy quark field. Note that using $p^+=m_H v^+$, one has the relation $x_{i,0}=\nu_i/m_H$. The corresponding matrix element up to ${\cal O}(\alpha_s)$ can be writen as
\begin{align}
&\langle0|{\cal O}_{h}(\omega_{1},\omega_{2})|{\overline {u(k_{1})d(k_{2})}}Q(p_{Q})\rangle\nonumber\\
    =&S^{(0)}\left[\delta(\omega_{1}-\nu_1)\delta(\omega_{1}-\nu_2)+\frac{\alpha_{s}C_{F}}{4\pi}N^{(1)}(\omega_{1},\omega_{2};\nu_1,\nu_2)\right].\label{eq:bHQETmatrix1loop}
\end{align}
Similarly to $M^{(1)}$, $N^{(1)}$  contains the same one-loop diagrams as shown in Fig.~\ref{fig:LCDAmatching}, ${\cal O}(\alpha_s)$ corrections to the light and residues of heavy quark field $h_n$, as well as the renormalization factor canceling  UV divergence.

The jet function up to ${\cal O}(\alpha_s)$ can be writen as:
\begin{align}
&{\cal J}_{\rm peak}(x_{1},x_{2};\omega_{1},\omega_{2})\nonumber\\
=&\ \theta\left(m_{H}-\omega_{1}-\omega_{2}\right)\left[\delta\left(x_{1}-\frac{\omega_{1}}{m_{H}}\right)\delta\left(x_{2}-\frac{\omega_{2}}{m_{H}}\right)+\frac{\alpha_{s}C_{F}}{4\pi}{\cal J}^{(1)}(x_{1},x_{2};\omega_{1},\omega_{2})\right]. \label{eq:Jpeakto1loopa}
\end{align}
Inserting Eq.~(\ref{eq:SCETmatrix1loop}), Eq.~(\ref{eq:bHQETmatrix1loop}) and Eq.~(\ref{eq:Jpeakto1loopa}) into Eq.~(\ref{eq:OcMatchOh}), one obtains the expression of ${\cal J}^{(1)}$
\begin{align}
    {\cal J}^{(1)}(x_{1},x_{2};\omega_{1},\omega_{2})= & M^{(1)}(x_{1},x_{2};\frac{\omega_{1}}{m_{H}},\frac{\omega_{2}}{m_{H}})-m_{H}^{2}N^{(1)}(x_{1}m_{H},x_{2}m_{H};\omega_{1},\omega_{2}).\label{eq:J1match}
\end{align}
From Eq.~\eqref{eq:J1match} one observes that all light‑quark self‑interaction diagrams: $W_{uu}$, $W_{dd}$, $W_{ud}$, and $W_{du}$ do not contribute to the matching, because their amplitudes are identical on the SCET and bHQET sides. Therefore, practically we will not include the light‑quark self‑interaction diagrams as well as the light quark residues into the calculation of $M^{(1)}$ and $N^{(1)}$. As a result, the diagrams we have to consider are
\begin{align}  W_{Q}^{c,h}&=W_{QQ}^{c,h}+W_{Qd}^{c,h}+W_{Qu}^{c,h},\nonumber\\
    W_{q}^{c,h}&=W_{uQ}^{c,h}+W_{dQ}^{c,h},\nonumber\\
    V_{Q}^{c,h}&=V_{Qu}^{c,h}+V_{Qd}^{c,h}.\label{eq:relevantDiagrams}
\end{align}
The superscripts $c$ and $h$ distinguish the same diagram calculated in SCET or in bHQET, respectively.
Then the relevant contribution of $M^{(1)}$ and $N^{(1)}$ to the matching read as
\begin{align}
    M^{(1)}(x_{1},x_{2};x_{1,0},x_{2,0})=&\left[W_{Q}^c+W_{q}^c+V_{Q}^c\right](x_{1},x_{2};x_{1,0},x_{2,0})+\frac{1}{2}Z_Q^{(1)}\delta(x_{1}-x_{1,0})\delta(x_{2}-x_{2,0})\nonumber\\
    &+Z_{\rm SCET}^{(1)}(x_{1},x_{2};x_{1,0},x_{2,0}),\\
    N^{(1)}(\omega_{1},\omega_{2};\nu_1,\nu_2)=&\left[W_{Q}^h+W_{q}^h+V_{Q}^h\right](\omega_{1},\omega_{2};\nu_1,\nu_2)+\frac{1}{2}Z_h^{(1)}\delta(\omega_{1}-\nu_{1})\delta(\omega_{2}-\nu_{2})\nonumber\\
    &+Z_{\rm bHQET}^{(1)}(\omega_{1},\omega_{2};\nu_1,\nu_2),\label{eq:M1N1diagrams}
\end{align}
where $Z_Q^{(1)}$ and $Z_h^{(1)}$ denotes the $\mathcal{O}(\alpha_s)$ correction to the residue of heavy‑quark field in massive SCET and bHQET, respectively:
\begin{align}
Z_{Q}^{(1)} =-\frac{3}{\epsilon}-3{\rm ln}\frac{\mu^{2}}{m_{Q}^{2}}-4,~~~~~~~~~~
Z_{h}^{(1)} = 0.\label{eq:quarkResidues}
\end{align}
Note that $Z_h^{(1)}$ vanishes due to scaleless integration.  $Z_{\mathrm{SCET}}^{(1)}$ and $Z_{\mathrm{bHQET}}^{(1)}$ are the renormalization factor that cancels all the UV divergences in $W_{Q}^{c,h} + W_{q}^{c,h} + V_{Q}^{c,h}$ and $Z_{Q,h}^{(1)}$.

In Sec.~\ref{sec:SCETcalculation} and Sec.~\ref{sec:bHQETcalculation}, we will present detailed calculation of the amplitudes $W_{Q}^{c}, W_{q}^{c},  V_{Q}^{c}$ and $W_{Q}^{h}, W_{q}^{h}, V_{Q}^{h}$. It will be found that $W_{q}^{c}=W_{q}^{h}=0$ due to scaleless integration. In the peak region, the calculation of $W_{Q}^{c}$ and $V_{Q}^{c}$ can be significantly simplified using the method of regions. As argued by Ref.~\cite{Beneke:2023nmj}, the integration of loop momentum $q$ can be separated into integrations in the hard-collinear region $q_{\rm hc} \sim (1,b,b^2) Q$ and the soft-collinear region $q_{\rm sc} \sim (1,b,b^2) \Lambda_{\rm QCD}Q/m_Q$ , namely
\begin{align}
    W_{Q}^{c}&=W_{Q,\rm hc}^{c}+W_{Q,\rm sc}^{c},\nonumber\\
    V_{Q}^{c}&=V_{Q,\rm hc}^{c}+V_{Q,\rm sc}^{c}.
\end{align}
As will be demonstrated in Sec.~\ref{sec:jetfunction}, in the peak region
\begin{align}
W_{Qu,\mathrm{hc}}^{c} = W_{Qd,\mathrm{hc}}^{c} = V_{Qu,\mathrm{hc}}^{c} = V_{Qd,\mathrm{hc}}^{c} = 0\label{eq:SCEThcVanish}
\end{align}
implying that all diagrams containing a gluon linking the light‑quark and heavy‑quark sectors yield vanishing hard‑collinear amplitudes. Furthermore, the soft‑collinear amplitudes in SCET are identical to their bHQET counterparts:
\begin{align}
    W_{Q,\rm sc}^{c}(x_{1},x_{2};\omega_{1},\omega_{2})&=m_H^2 W_{Q,\rm sc}^{h}(x_{1}m_{H},x_{2}m_{H};\omega_{1},\omega_{2}),\nonumber\\
    V_{Q,\rm sc}^{c}(x_{1},x_{2};\omega_{1},\omega_{2})&=m_H^2 V_{Q,\rm sc}^{h}(x_{1}m_{H},x_{2}m_{H};\omega_{1},\omega_{2}).\label{eq:SCETsoftbHQET}
\end{align}
This identity will also be proved in Sec.~\ref{sec:jetfunction}.
Using these simplified amplitudes and Eq.~(\ref{eq:J1match}), one can derive the jet function as
\begin{align}
    {\cal J}^{(1)}(x_{1},x_{2};\omega_{1},\omega_{2})=& W_{QQ,\mathrm{hc}}^{c}\left(x_{1},x_{2};\frac{\omega_{1}}{m_{H}},\frac{\omega_{2}}{m_{H}}\right)+\frac{1}{2}Z_Q^{(1)}\delta\left(x_{1}-\frac{\omega_{1}}{m_{H}}\right)\delta\left(x_{2}-\frac{\omega_{2}}{m_{H}}\right)\nonumber\\
    &+Z_{\rm SCET}^{(1),\rm peak}\left(x_{1},x_{2};\frac{\omega_{1}}{m_{H}},\frac{\omega_{2}}{m_{H}}\right)-m_H^2 Z_{\rm bHQET}^{(1)}\left(x_{1}m_{H},x_{2}m_{H};\omega_{1},\omega_{2}\right),\label{eq:jetMatchingFormula}
\end{align}
where $Z_{\mathrm{SCET}}^{(1),\mathrm{peak}}$ denotes the leading‑power expansion of $Z_{\mathrm{SCET}}^{(1)}$ in the small $x_i \sim x_{i,0}$ limit, taken in the peak‑region regime $x_i \sim x_{i,0} \sim \Lambda_{\mathrm{QCD}}/m_Q$.
Therefore, employing the method of regions, only the hard‑collinear amplitude of the $W_{QQ}$ diagram needs to be computed to extract the jet function in the peak region. It should be noted, however, that the UV part of all the amplitudes $W_{Q}^{c}$, $W_{q}^{c}$, and $V_{Q}^{c}$ still has to be included in the calculation of $Z_{\mathrm{SCET}}^{(1)}$.  

\section{SCET amplitudes}\label{sec:SCETcalculation}

This section focus on the calculation of $W_{Q}^{c}$, $W_{q}^{c}$, and $V_{Q}^{c}$ at leading power in SCET.
At leading power, the SCET calculation reduces to the QCD result by retaining only the large component of the heavy‑quark spinor
\begin{align}
    Q(p_Q)=\frac{\bar{\slashed n}\slashed n}{2}Q(p_Q).\label{eq:QLargeComponent}
\end{align}
The tree level amplitude $S^{(0)}$ can be easily obtained by calculating the tree level matrix element defined in Eq.~(\ref{eq:SCETmatrix1loop}):
\begin{align}
    W^c_{\rm tree}=&\delta(x_{1}-x_{1,0})\delta(x_{2}-x_{2,0})u^{T}(k_{1})\Gamma d(k_{2})Q(p_{Q}).\label{eq:SCETtreeMatrix}
\end{align}
The spin averaging defined in Eq.~(\ref{eq:spinaverage}) leads to simplification on the light quark spinors:
\begin{align}
u^{T}(k_{1})\Gamma d(k_{2}) & ={\rm tr}\left[d(k_{2}) u^{T}(k_{1})\Gamma\right]\nonumber\\
 & =\frac{1}{\sqrt{2}}{\rm tr}\left[d_{\downarrow}(k_{2}) u_{\uparrow}^{T}(k_{1})\Gamma-d_{\uparrow}(k_{2}) u_{\downarrow}^{T}(k_{1})\Gamma\right].\label{eq:spinAverage1}
\end{align}
Using the momentum fractions $k_i^+=x_{i,0}p^+$ and the identity:
\begin{align}
 & \frac{1}{\sqrt{2}}\left[d_{\downarrow}(x_{2,0}p^{+}\bar{n})u_{\uparrow}^{T}(x_{1,0}p^{+}\bar{n})\Gamma-d_{\uparrow}(x_{2,0}p^{+}\bar{n})u_{\downarrow}^{T}(x_{1,0}p^{+}\bar{n})\Gamma\right]=c_1 \frac{1}{2}p^{+}\bar{\slashed n}C\gamma_{5}\Gamma, \label{eq:spinAverage2}
\end{align}
one obtains
\begin{align}
    W^c_{\rm tree}=&\delta(x_{1}-x_{1,0})\delta(x_{2}-x_{2,0})c_1 S^{(0)}.\label{eq:SCETtreeMatrix1}
\end{align}
Here $c_1 = -\sqrt{2 x_{1,0} x_{2,0}}$ and $S^{(0)}=-2p^{+}Q(p_{Q})$. Because $c_1$ appears with the same value in both the tree‑level and one‑loop matrix elements, and is identical in the SCET and bHQET calculations, this factor can be neglected or simply set to $c_1 = 1$.

According to Eq.~(\ref{eq:jetMatchingFormula}), only $W_{QQ}^c$ needs to be calculated in full. In the hard‑collinear region, where the loop momentum scales as $q_{\mathrm{hc}} \sim (1,b,b^{2}) Q$, the amplitude of $W_{QQ,\rm hc}^c$ reads
\begin{align}
    \frac{\alpha_{s}C_{F}}{4\pi}W_{QQ,\rm hc}^c=& \ C_{F}ig_{s}^{2}\int\frac{D^{d}q}{(2\pi)^{4}}\delta(x_{1}-x_{1,0})\delta(x_{2}-x_{2,0})\frac{1}{q\cdot n-i\varepsilon}\\
 & \times\frac{1}{q^2+i\varepsilon}\frac{1}{(p_{Q}-q)^{2}-m_{Q}^{2}+i\varepsilon}u^{T}(k_{1})\Gamma d(k_{2})(\slashed p_{Q}-\slashed q+m_{Q})\slashed n\ Q(p_{Q}),\label{eq:WQQamplitude0}
\end{align}
where an extra factor has been included in the integration measure to implement the $\overline{\mathrm{MS}}$ renormalization scheme:
\begin{align}
    D^{d}q=\left(\frac{\mu^{2}}{(4\pi)e^{-\gamma_{E}}}\right)^{\epsilon}d^{d}q.
\end{align}
At leading power the numerator can be simplfied by Eq.~(\ref{eq:QLargeComponent}) as
\begin{align}
  u^{T}(k_{1})\Gamma d(k_{2})(\slashed p_{Q}-\slashed q+m_{Q})\slashed n\ Q(p_{Q})= (2p_{Q}^{+}-2q^{+})u^{T}(k_{1})\Gamma d(k_{2})Q(p_{Q}).\label{eq:WQQnumerate}
\end{align}
Performing the spin average as shown in Eqs.~\eqref{eq:spinAverage1} and \eqref{eq:spinAverage2}, the numerator can be further simplified to $(2p_{Q}^{+}-2q^{+})S^{(0)}$. After the loop momentum integration, the $W_{QQ}$ amplitude reads as
\begin{align}
    W_{QQ,\rm hc}^c=\delta(x_{1}-x_{1,0})\delta(x_{2}-x_{2,0})\left[\frac{1}{\epsilon^{2}}+\frac{1}{\epsilon}{\rm ln}\frac{\mu^{2}}{m_{Q}^{2}}+\frac{2}{\epsilon}+\frac{1}{2}{\rm ln}^{2}\frac{\mu^{2}}{m_{Q}^{2}}+2{\rm ln}\frac{\mu^{2}}{m_{Q}^{2}}+\frac{\pi^{2}}{12}+4\right].\label{eq:WQQhcResult}
\end{align}

Although only the hard‑collinear part of $W_{QQ}^c$ is required by the matching formula in Eq.~\eqref{eq:jetMatchingFormula},  in the Appendix~\ref{sec:CompleteAmpSCET} we list the full amplitudes of all the diagrams shown in Fig. for completeness. Nevertheless, the UV part of each diagram is still necessary to determine $Z_{\mathrm{SCET}}^{(1)}$. The UV part of each diagram are
\begin{align}
    W_{QQ,UV}^{c}= & -\delta(x_{1}-x_{1,0})\delta(x_{2}-x_{2,0}) \frac{2}{\epsilon}\int_{0}^{1}d\lambda\frac{1-\lambda}{\lambda},\label{eq:WQQUV}\\
    W_{Qu,UV}^{c}= & \frac{1}{2}\delta(x_{2}-x_{2,0}) \frac{2}{\epsilon}\left[\theta(\bar{x}_{2,0}-x_{1})\theta(x_{1}-x_{1,0})\frac{\bar{x}_{2,0}-x_{1}}{(x_{1}-x_{1,0})x_{3,0}}\right]_{x_{1}+}\nonumber\\
 & +\frac{1}{2}\delta(x_{1}-x_{1,0})\delta(x_{2}-x_{2,0}) \frac{2}{\epsilon}\int_{0}^{1}d\lambda\frac{1-\lambda}{\lambda},\label{eq:WQuUV}\\
 W_{uQ,UV}^{c}= & \frac{1}{2}\delta(x_{2}-x_{2,0}) \frac{2}{\epsilon}\left[\frac{\theta(x_{1,0}-x_{1})}{x_{1,0}-x_{1}}\frac{x_{1}}{x_{1,0}}\right]_{x_{1}+}\nonumber\\
 & +\frac{1}{2}\delta(x_{1,0}-x_{1})\delta(x_{2}-x_{2,0}) \frac{2}{\epsilon}\int_{0}^{1}d\lambda\ \frac{1-\lambda}{\lambda},\label{eq:WuQUV}\\
 W_{uu,UV}^{c}= & -\delta(x_{1}-x_{1,0})\delta(x_{2}-x_{2,0}) \frac{2}{\epsilon}\int_{0}^{1}d\lambda\frac{1-\lambda}{\lambda},\\
W_{ud,UV}^{c}= & \frac{1}{2}\delta(x_{3}-x_{3,0}) \frac{2}{\epsilon}\left[\frac{\theta(x_{1,0}-x_{1})}{x_{1,0}-x_{1}}\frac{x_{1}}{x_{1,0}}\right]_{x_{1}+}\nonumber\\
 & +\frac{1}{2}\delta(x_{1,0}-x_{1})\delta(x_{2}-x_{2,0}) \frac{2}{\epsilon}\int_{0}^{1}d\lambda\ \frac{1-\lambda}{\lambda},\\
 V_{Qu,UV}^{c}= & \frac{1}{2}\delta(x_{2}-x_{2,0})\theta(x_{1,0}-x_{1}) \frac{x_{1}}{x_{1,0}(1-x_{2})}\frac{1}{\epsilon}\nonumber\\
 & \frac{1}{2}\delta(x_{2}-x_{2,0})\theta(x_{1,0}+x_{3,0}-x_{1})\theta(x_{1}-x_{1,0}) \frac{1}{x_{1,0}}\left[\frac{x_{1}}{x_{1,0}+x_{3,0}}-\frac{x_{1}-x_{1,0}}{x_{3,0}}\right]\frac{1}{\epsilon},\label{eq:VudUV}\\
 V_{ud,UV}^{c}= & \delta(x_{3}-x_{3,0}) \left[\frac{x_{1}\theta(x_{2}-x_{2,0})}{x_{1,0}(x_{1,0}+x_{2,0})}+\frac{x_{2}\theta(x_{1}-x_{1,0})}{x_{2,0}(x_{1,0}+x_{2,0})}\right]\frac{1}{\epsilon},\label{eq:VudUV}
\end{align}
where $\bar{x}_{i,0}=1-x_{i,0}$, and the plus function is defined, for a function $f(x,x_0)$ that is singular at $x_0$, by
\begin{align}
    \left[f(x,x_0)\right]_{x+}=f(x,x_0)-\delta(x-x_0)\int_{-\infty}^{\infty }dt\ f(t,x_0).
\end{align}
The remaining UV amplitudes can be obtained via the following replacements:
\begin{align}
&\left(W_{Qd,UV}^{c},~W_{dQ,UV}^{c},~W_{dd,UV}^{c},~W_{du,UV}^{c},~V_{Qd,UV}^{c}\right)\nonumber\\
=&\left(W_{Qu,UV}^{c},~W_{uQ,UV}^{c},~W_{uu,UV}^{c},~W_{ud,UV}^{c},~V_{Qu,UV}^{c}\right)_{x_1\leftrightarrow x_2,~x_{1,0}\leftrightarrow x_{2,0}}.
\end{align}
The divergent integration over $\lambda$ in the $W$ diagrams stems from the rapidity divergences. Such divergences, however, cancel between virtual and real diagrams. As an example, the rapidity divergences in $W_{QQ,\mathrm{UV}}^{c}$ are cancelled by those in $W_{Qu,\mathrm{UV}}^{c}+W_{Qd,\mathrm{UV}}^{c}$. On the other hand, we note that the amplitudes $W_{uu,\mathrm{UV}}^{c}$, $W_{dd,\mathrm{UV}}^{c}$, $W_{ud,\mathrm{UV}}^{c}$, $W_{du,\mathrm{UV}}^{c}$ and $V_{ud,\mathrm{UV}}^{c}$ are identical in the SCET and bHQET calculations, and therefore do not contribute to the matching for ${\cal J}^{(1)}$. Thus the UV part of the required amplitudes are
\begin{flalign}
  &W_{Q,UV}^{c}(x_{1},x_{2};x_{1,0},x_{2,0})= \left(W_{QQ,UV}^{c}+W_{Qu,UV}^{c}+W_{Qd,UV}^{c}\right)(x_{1},x_{2};x_{1,0},x_{2,0})&\nonumber\\
= & \frac{1}{2} \delta(x_{1}-x_{1,0}) \frac{2}{\epsilon}\left[\theta(\bar{x}_{1,0}-x_{2})\theta(x_{2}-x_{2,0})\frac{\bar{x}_{1,0}-x_{2}}{(x_{2}-x_{2,0})(\bar{x}_{1,0}-x_{2,0})}\right]_{x_{2,0}+}&\nonumber\\
 & +\frac{1}{2} \delta(x_{2}-x_{2,0}) \frac{2}{\epsilon}\left[\theta(\bar{x}_{2,0}-x_{1})\theta(x_{1}-x_{1,0})\frac{\bar{x}_{2,0}-x_{1}}{(x_{1}-x_{1,0})(\bar{x}_{2,0}-x_{1,0})}\right]_{x_{1,0}+}&\nonumber\\
 & +  \delta(x_{1}-x_{1,0})\delta(x_{2}-x_{2,0})\frac{2}{\epsilon}\left(1+{\rm ln}x_{3,0}-\frac{1}{2}{\rm ln}\bar{x}_{1,0}-\frac{1}{2}{\rm ln}\bar{x}_{2,0}\right),& \label{eq:WQexpression} 
\end{flalign}
\begin{flalign}
    &W_{q,UV}^{c}(x_{1},x_{2};x_{1,0},x_{2,0})= \left(W_{uQ,UV}^{c}+W_{dQ,UV}^{c}\right)(x_{1},x_{2};x_{1,0},x_{2,0})&\nonumber\\
 =& \frac{1}{2} \delta(x_{2}-x_{2,0}) \frac{2}{\epsilon}\left[\frac{\theta(x_{1,0}-x_{1})}{x_{1,0}-x_{1}}\frac{x_{1}}{x_{1,0}}\right]_{x_{1,0}+}+\frac{1}{2} \delta(x_{1}-x_{1,0}) \frac{2}{\epsilon}\left[\frac{\theta(x_{2,0}-x_{2})}{x_{2,0}-x_{2}}\frac{x_{2}}{x_{2,0}}\right]_{x_{2,0}+}&\nonumber\\
 & +\frac{1}{2} \delta(x_{1}-x_{1,0})\delta(x_{2}-x_{2,0}) \frac{2}{\epsilon}\left(2+{\rm ln}x_{1,0}+{\rm ln}x_{2,0}\right)&\nonumber\\
 &+ \delta(x_{1}-x_{1,0})\delta(x_{2}-x_{2,0}) \frac{2}{\epsilon}\int_{0}^{1}d\lambda\frac{1-\lambda}{\lambda},& \label{eq:Wqexpression}
\end{flalign}
\begin{flalign}
    &V_{Q,UV}^{c}(x_{1},x_{2};x_{1,0},x_{2,0})= \left(V_{Qu,UV}^{c}+V_{Qd,UV}^{c}\right)(x_{1},x_{2};x_{1,0},x_{2,0})&\nonumber\\
 =& \frac{1}{2} \delta(x_{1}-x_{1,0}) \frac{1}{\epsilon}\left[\theta(x_{2,0}-x_{2})\frac{x_{2}}{x_{2,0}(1-x_{1})}\right]_{x_{2,0}+}&\nonumber\\
 & +\frac{1}{2} \delta(x_{1}-x_{1,0}) \frac{1}{\epsilon}\left[\theta(\bar{x}_{1,0}-x_{2})\theta(x_{2}-x_{2,0})\frac{\bar{x}_{1,0}-x_{2}}{\bar{x}_{1,0}(\bar{x}_{1,0}-x_{2,0})}\right]_{x_{2,0}+}&\nonumber\\
 & +\frac{1}{2} \delta(x_{2}-x_{2,0}) \frac{1}{\epsilon}\left[\theta(x_{1,0}-x_{1})\frac{x_{1}}{x_{1,0}(1-x_{2})}\right]_{x_{1,0+}}&\nonumber\\
 & +\frac{1}{2} \delta(x_{2}-x_{2,0}) \frac{1}{\epsilon}\left[\theta(\bar{x}_{2,0}-x_{1})\theta(x_{1}-x_{1,0})\frac{\bar{x}_{2,0}-x_{1}}{\bar{x}_{2,0}(\bar{x}_{2,0}-x_{1,0})}\right]_{x_{1,0+}}&\nonumber\\
 & -\frac{1}{2} \delta(x_{1}-x_{1,0})\delta(x_{2,0}-x_{2}) \frac{1}{\epsilon}\left[\frac{x_{2}{\rm ln}x_{2}}{1-x_{1}}+\frac{x_{1}{\rm ln}x_{1}}{1-x_{2}}+\frac{\bar{x}_{1,0}-x_{2}}{\bar{x}_{1,0}}{\rm ln}\frac{x_{3,0}}{\bar{x}_{1,0}}+\frac{\bar{x}_{2,0}-x_{1}}{\bar{x}_{2,0}}{\rm ln}\frac{x_{3,0}}{\bar{x}_{2,0}}\right].& \label{eq:VQexpression}
\end{flalign}
For simplicity in the matching procedure, now the UV amplitudes are expressed using plus functions of the external momentum fractions $x_{i,0}$ instead of $x_i$. The last term in $W_{q,\mathrm{UV}}^{c}$ is a rapidity divergence that would be cancelled by the corresponding rapidity divergences in $W_{uu,\mathrm{UV}}^{c}$, $W_{dd,\mathrm{UV}}^{c}$, $W_{ud,\mathrm{UV}}^{c}$, and $W_{du,\mathrm{UV}}^{c}$. Since these four light‑quark self‑interaction diagrams are identical to their bHQET counterparts, they do not contribute to the matching and are not considered here. Consequently, the rapidity divergence in $W_{q,\mathrm{UV}}^{c}$ remains uncancelled. However, an identical divergence appears in the bHQET amplitude $W_{q,\mathrm{UV}}^{h}$, so it also cancels in the matching and can be neglected for simplicity.

Now we take the peak region limit $x_{i}\sim x_{i,0}\sim\Lambda_{\rm QCD}/m_Q\ll1$ for the UV amplitudes given above. In this limit, using the approximation: $x_{3,0}=1-x_{1,0}-x_{2,0}\approx 1$, $\bar{x}_{i,0}\approx 1$ and $\theta(\bar{x}_{1,0}-x_{2})=\theta(\bar{x}_{2,0}-x_{1})\approx1$, one can reduce the required UV amplitudes as
\begin{align}
& W_{Q,UV}^{c, \rm peak}(x_{1},x_{2};x_{1,0},x_{2,0})\nonumber\\
= & \frac{1}{2}\delta(x_{1}-x_{1,0})\frac{2}{\epsilon}\left[\frac{\theta(x_{2}-x_{2,0})}{x_{2}-x_{2,0}}\right]_{x_{2,0}+}+\frac{1}{2}\delta(x_{2}-x_{2,0})\frac{2}{\epsilon}\left[\frac{\theta(x_{1}-x_{1,0})}{x_{1}-x_{1,0}}\right]_{x_{1,0}+}\nonumber\\
 & +\delta(x_{1}-x_{1,0})\delta(x_{2}-x_{2,0})\frac{2}{\epsilon},\\
& W_{q,UV}^{c, \rm peak}(x_{1},x_{2};x_{1,0},x_{2,0})\nonumber\\
 = & \frac{1}{2}\delta(x_{2}-x_{2,0})\frac{2}{\epsilon}\left[\frac{\theta(x_{1,0}-x_{1})}{x_{1,0}-x_{1}}\frac{x_{1}}{x_{1,0}}\right]_{x_{1,0}+} +\frac{1}{2}\delta(x_{1}-x_{1,0})\frac{2}{\epsilon}\left[\frac{\theta(x_{2,0}-x_{2})}{x_{2,0}-x_{2}}\frac{x_{2}}{x_{2,0}}\right]_{x_{2,0}+}\nonumber\\
 & +\frac{1}{2}\delta(x_{1}-x_{1,0})\delta(x_{2}-x_{2,0})\frac{2}{\epsilon}\left(2+{\rm ln}x_{1,0}+{\rm ln}x_{2,0}\right)\nonumber\\
 & +\delta(x_{1}-x_{1,0})\delta(x_{2}-x_{2,0})\frac{2}{\epsilon}\int_{0}^{1}d\lambda\frac{1-\lambda}{\lambda},\\
& V_{q,UV}^{c, \rm peak}(x_{1},x_{2};x_{1,0},x_{2,0})\nonumber\\
 = & \frac{1}{2}\delta(x_{1}-x_{1,0})\frac{1}{\epsilon}\left[\theta(x_{2,0}-x_{2})\frac{x_{2}}{x_{2,0}}+\theta(x_{2}-x_{2,0})\right]_{x_{2,0}+}\nonumber\\
 & +\frac{1}{2}\delta(x_{2}-x_{2,0})\frac{1}{\epsilon}\left[\theta(x_{1,0}-x_{1})\frac{x_{1}}{x_{1,0}}+\theta(x_{1}-x_{1,0})\right]_{x_{1,0+}}.\label{eq:UVatpeak}
\end{align}
The renormalization factor corresponding to the UV amplitudes given in Eq.~(\ref{eq:UVatpeak}) reads as
\begin{align}
 & Z_{SCET}^{(1), \rm peak}(x_{1},x_{2};x_{1,0},x_{2,0})\nonumber \\
= & -\left[W_{Q,UV}^{c,\rm peak}+W_{q,UV}^{c,\rm peak}+V_{Q,UV}^{c,\rm peak}\right](x_{1},x_{2};x_{1,0},x_{2,0})-\delta\left(x_{1}-x_{1,0}\right)\delta\left(x_{2}-x_{2,0}\right)\frac{1}{2}Z_{Q}^{UV}\frac{1}{\epsilon}\nonumber \\
= & -\frac{1}{2}\delta(x_{1}-x_{1,0})\frac{2}{\epsilon}\left[\frac{\theta(x_{2}-x_{2,0})}{x_{2}-x_{2,0}}+\frac{x_{2}}{x_{2,0}}\frac{\theta(x_{2,0}-x_{2})}{x_{2,0}-x_{2}}\right]_{x_{2,0}+}\nonumber \\
 & -\frac{1}{2}\delta(x_{2}-x_{2,0})\frac{2}{\epsilon}\left[\frac{\theta(x_{1}-x_{1,0})}{x_{1}-x_{1,0}}+\frac{x_{1}}{x_{1,0}}\frac{\theta(x_{1,0}-x_{1})}{x_{1,0}-x_{1}}\right]_{x_{1,0}+}\nonumber \\
 & -\frac{1}{2}\delta(x_{1}-x_{1,0})\delta(x_{2}-x_{2,0})\frac{2}{\epsilon}\left(4+\frac{1}{2}Z_{Q}^{UV}+{\rm ln}x_{1,0}+{\rm ln}x_{2,0}\right)\nonumber\\
 & -\delta(x_{1}-x_{1,0})\delta(x_{2}-x_{2,0})\frac{2}{\epsilon}\int_{0}^{1}d\lambda\frac{1-\lambda}{\lambda},\label{eq:ZSCET}
\end{align}
where $Z_{Q}^{UV}=-1$ is the UV part of the heavy quark pole residue at SCET.

\section{bHQET amplitudes}\label{sec:bHQETcalculation}

This section focus on the calculate of $W_{Q}^{h}$, $W_{q}^{h}$, and $V_{Q}^{h}$ at leading power in bHQET. The tree‑level matrix element in bHQET can be derived in the same way as in SCET. Employing the spin‑averaging procedure yields
\begin{align}
    W_{\rm tree}^c=\delta\left(\omega_{1}-\nu_{1}\right)\delta\left(\omega_{2}-\nu_{2}\right)\tilde{c}_{1}S^{(0)},
\end{align}
where $\tilde{c}_{1}=-\sqrt{2\omega_{1}\omega_{2}}/m_{H}=c_1$, and $S^{(0)}=-2p^{+}Q(p_{Q})$ is the same as that of SCET.

The complete bHQET amplitudes of all the diagram in Fig. are given in appendix. According to Eq.~(\ref{eq:jetMatchingFormula}), only the UV part of the amplitudes at bHQET are necessary. The UV parts of each diagram read as
\begin{flalign}
&W_{Q,UV}^{h}(\omega_{1},\omega_{2};\nu_1,\nu_2)= \left(W_{QQ,UV}^{h}+W_{Qu,UV}^{h}+W_{Qd,UV}^{h}\right)(\omega_{1},\omega_{2};\nu_1,\nu_2)&\nonumber\\
    = & \frac{1}{2} \delta(\omega_{1}-\nu_{1})\frac{2\omega_{2}}{\epsilon}\left[\frac{\theta(\omega_{2}-\nu_{2})}{\omega_{2}(\omega_{2}-\nu_{2})}\right]_{\omega_{2}+}-\frac{1}{2}\delta\left(\omega_{1}-\nu_{1}\right)\delta\left(\omega_{2}-\nu_{2}\right) \left[\frac{1}{\epsilon^{2}}+\frac{1}{\epsilon}{\rm ln}\frac{\mu^{2}}{\omega_{2}^{2}}\right]&\nonumber \\
 & +\frac{1}{2} \delta(\omega_{2}-\nu_{2})\frac{2\omega_{1}}{\epsilon}\left[\frac{\theta(\omega_{1}-\nu_{1})}{\omega_{1}(\omega_{1}-\nu_{1})}\right]_{\omega_{1}+}-\frac{1}{2}\delta\left(\omega_{1}-\nu_{1}\right)\delta\left(\omega_{2}-\nu_{2}\right) \left[\frac{1}{\epsilon^{2}}+\frac{1}{\epsilon}{\rm ln}\frac{\mu^{2}}{\omega_{1}^{2}}\right],&
 \end{flalign}
 \begin{flalign}
 &W_{q,UV}^{h}(\omega_{1},\omega_{2};\nu_1,\nu_2)= \left(W_{uQ,UV}^{h}+W_{dQ,UV}^{h}\right)(\omega_{1},\omega_{2};\nu_1,\nu_2)&\nonumber\\
 = & \frac{1}{2}\delta(\omega_{2}-\nu_{2})\frac{2\omega_{1}}{\epsilon}\left[\frac{\theta(\nu_{1}-\omega_{1})}{\nu_{1}(\nu_{1}-\omega_{1})}\right]_{\omega_{1}+} +\frac{1}{2}\delta(\omega_{1}-\nu_{1})\frac{2\omega_{2}}{\epsilon}\left[\frac{\theta(\nu_{2}-\omega_{2})}{\nu_{2}(\nu_{2}-\omega_{2})}\right]_{\omega_{2}+}&\nonumber \\
 & +\delta(\omega_{1}-\nu_{1})\delta(\omega_{2}-\nu_{2})\frac{2}{\epsilon}\left[1+\int_{0}^{1}d\lambda\frac{1-\lambda}{\lambda}\right],&
 \end{flalign}
 \begin{flalign}
 & V_{uQ,UV}^{h}=V_{dQ,UV}^{h}=0.\label{eq:bHQETUV}
\end{flalign}
The renormalization factor corresponding to the UV amplitudes in Eq.~(\ref{eq:bHQETUV}) is
\begin{align}
 & Z_{bHQET}^{(1)}(\omega_{1},\omega_{2};\nu_{1},\nu_{2})\nonumber \\
=& -\left[W_{Q,UV}^{h,\rm peak}+W_{q,UV}^{h,\rm peak}+V_{Q,UV}^{h,\rm peak}\right](\omega_{1},\omega_{2};\nu_{1},\nu_{2})-\delta\left(\omega_{1}-\nu_{1}\right)\delta\left(\omega_{2}-\nu_{2}\right)\frac{1}{2}Z_{h}^{UV}\frac{1}{\epsilon}\nonumber \\ 
= & -\frac{1}{2}\delta(\omega_{1}-\nu_{1})\frac{2}{\epsilon}\left[\frac{\theta(\omega_{2}-\nu_{2})}{\omega_{2}-\nu_{2}}+\frac{\omega_{2}\theta(\nu_{2}-\omega_{2})}{\nu_{2}(\nu_{2}-\omega_{2})}\right]_{\nu_{2}+}\nonumber\\
&-\frac{1}{2}\delta(\omega_{2}-\nu_{2})\frac{2}{\epsilon}\left[\frac{\theta(\omega_{1}-\nu_{1})}{\omega_{1}-\nu_{1}}+\frac{\omega_{1}\theta(\nu_{1}-\omega_{1})}{\nu_{1}(\nu_{1}-\omega_{1})}\right]_{\nu_{1}+}\nonumber\\
& +\frac{1}{2}\delta\left(\omega_{1}-\nu_{1}\right)\delta\left(\omega_{2}-\nu_{2}\right)\left[\frac{2}{\epsilon^{2}}-\frac{4}{\epsilon}-Z_{h}^{UV}\frac{1}{\epsilon}+\frac{1}{\epsilon}{\rm ln}\frac{\mu^{2}}{\nu_{1}^{2}}+\frac{1}{\epsilon}{\rm ln}\frac{\mu^{2}}{\nu_{2}^{2}}\right]\nonumber \\
 & -\delta(\omega_{1}-\nu_{1})\delta(\omega_{2}-\nu_{2})S^{(0)}\frac{2}{\epsilon}\int_{0}^{1}d\lambda\frac{1-\lambda}{\lambda}.\label{eq:ZbHQET}
\end{align}
Here the result are written using plus functions of external momentums $\nu_{i}$ instead of $\omega_i$. The last term above is the rapidity divergence originating from $W_{q,\mathrm{UV}}^{h}$, which will cancel against the last term in Eq.~\eqref{eq:ZSCET} during the matching. $Z_{h}^{UV}=2$ is the UV part of the heavy quark pole residue at bHQET.

\section{The Jet Function}\label{sec:jetfunction}
Before carrying out the matching, we first establish two key results of the method-of-regions: the vanishing of hard‑collinear contributions stated in Eq.~\eqref{eq:SCEThcVanish}, and the equivalence between the soft‑collinear amplitudes in SCET and their bHQET counterparts, as expressed in Eq.~\eqref{eq:SCETsoftbHQET}. 

The full amplitude of $W_{QQ}^{c}$ is identical to its hard‑collinear expression given in Eq.~\eqref{eq:WQQamplitude0}. In the soft‑collinear region, where the loop momentum scales as $q_{\mathrm{sc}} \sim (1,b,b^{2}) \Lambda_{\mathrm{QCD}}Q/m_Q$, the corresponding amplitude is obtained by omitting the terms $\slashed{q}$ and $q^{2}$ from the numerator and denominator of Eq.~\eqref{eq:WQQamplitude0}, respectively:
\begin{align}
\frac{\alpha_sC_F}{4\pi}W_{QQ,\rm sc}^{c} & =C_{F}ig_{s}^{2}\int\frac{D^{d}q}{(2\pi)^{4}}\delta(x_{1}-x_{1,0})\delta(x_{2}-x_{2,0})\frac{1}{q\cdot n-i\varepsilon}\frac{1}{q^{2}+i\varepsilon}\frac{2p_{Q}^{+}}{-2p_{Q}\cdot q+i\varepsilon}\nonumber\\
 & =-C_{F}(n\cdot v)ig_{s}^{2}\delta(x_{1}-x_{1,0})\delta(x_{2}-x_{2,0})\int\frac{D^{d}q}{(2\pi)^{4}}\frac{1}{q\cdot n+i\varepsilon}\frac{1}{q^{2}+i\varepsilon}\frac{1}{v\cdot q+i\varepsilon}.
\end{align}
In bHQET, using the leading power Lagrangian in Eq.~(\ref{eq:bHQETLagran}), one can write the amplitude of the same diagram as
\begin{align}
\frac{\alpha_sC_F}{4\pi}W_{QQ,\rm sc}^{h}= & -C_{F}(n\cdot v)(ig_{s}^{2})\delta\left(\omega_{1}-\nu_{1}\right)\delta\left(\omega_{2}-\nu_{2}\right)\int\frac{D^{d}q}{(2\pi)^{4}}\frac{1}{q\cdot n+i\varepsilon}\frac{1}{q^{2}+i\varepsilon}\frac{1}{v\cdot q+i\varepsilon}.
\end{align}
It is obvious that these two expressions satisfy the relation given in Eq.~(\ref{eq:SCETsoftbHQET}). 

The full amplitude of $W_{Qd}^{c}$ is expressed as:
\begin{align}
\frac{\alpha_sC_F}{4\pi}W_{Qd}^{c}= & -\frac{C_{F}}{2}ig_{s}^{2}\int\frac{D^{d}q}{(2\pi)^{4}}\delta(x_{1}-x_{1,0})\delta\left(x_{2}-x_{2,0}+\frac{q^{+}}{p^{+}}\right)\frac{1}{(q\cdot n-i\varepsilon)}\nonumber\\
 & \times\frac{1}{q^2+i\varepsilon}\frac{1}{(p_{Q}-q)^{2}-m_{Q}^{2}+i\varepsilon}u^{T}(k_{1})\Gamma d(k_{2})(\slashed p_{Q}-\slashed q+m_{Q})\slashed n\ Q(p_{Q}),\label{eq:WQdcAmplitude}
\end{align}
which has exactly the same form of propagators as that of $W_{QQ}^c$ in Eq.~(\ref{eq:WQQamplitude0}). However, the loop momentum now enters inside a delta‑function $\delta\left(x_{2}-x_{2,0}+q^{+}/p^{+}\right)$ in the integrand. In the peak region, $x_{2} \sim x_{2,0} \sim \Lambda_{\mathrm{QCD}}/m_Q$, so that $x_{2} - x_{2,0} \sim \Lambda_{\mathrm{QCD}}/m_Q$, whereas the hard‑collinear momentum scales as $q_{\mathrm{hc}}^{+}/p^{+} \sim 1$. Consequently, this delta‑function forces the integral to vanish in the hard‑collinear region. Hence, in the peak region $W_{Qd}^{c}$ receives contributions only from the soft‑collinear region:
\begin{align}
    \frac{\alpha_sC_F}{4\pi}W_{Qd,\rm sc}^{c}=& \frac{C_{F}}{2}(n\cdot v)ig_{s}^{2}\delta(x_{1}-x_{1,0})\nonumber\\
    &\times\int\frac{D^{d}q}{(2\pi)^{4}}\delta\left(x_{2}-x_{2,0}+\frac{q^{+}}{p^{+}}\right)\frac{1}{q\cdot n+i\varepsilon}\frac{1}{q^{2}+i\varepsilon}\frac{1}{v\cdot q+i\varepsilon}.\label{eq:WQdcAmplitudesc}
\end{align}
On the other hand, the corresponding bHQET amplitude is
\begin{align}
    \frac{\alpha_sC_F}{4\pi}W_{Qd}^{h}=&\frac{C_{F}}{2}(n\cdot v)ig_{s}^{2}\delta(\omega_{1}-\nu_{1})\nonumber\\
    &\times\int\frac{D^{d}q}{(2\pi)^{4}}\delta\left(\omega_{2}-\nu_{2}+\frac{q^{+}}{v^{+}}\right)\frac{1}{q\cdot n+i\varepsilon}\frac{1}{q^{2}+i\varepsilon}\frac{1}{v\cdot q+i\varepsilon}.\label{eq:WQdcbHQET}
\end{align}
Clearly, Eq.~\eqref{eq:WQdcAmplitudesc} and Eq.~\eqref{eq:WQdcbHQET} also satisfy the relation stated in Eq.~\eqref{eq:SCETsoftbHQET}. The equivalence between $W_{Qu,\mathrm{sc}}^{c}$ and $W_{Qu}^{h}$ follows analogously.

Then we consider the amplitude of $V_{Qd}^{c}$ and $V_{Qu}^{c}$ in the method-of-regions. The complete amplitude of $V_{Qd}^{c}$ at leading power is
\begin{align}
    \frac{\alpha_sC_F}{4\pi}V_{Qd}^{c}=& \frac{C_{F}}{2}(p^{+})^{2}(ig_{s}^{2})\delta(x_{1}p^{+}-k_{1}^{+})\nonumber\\
    &\times\int\frac{D^{d}q}{(2\pi)^{4}}\frac{1}{q^{2}+i\varepsilon}\frac{1}{(q-k_{2})^{2}+i\varepsilon}\frac{\vec{q}_{\perp}^{2}+4p_{Q}^{-}(k_{2}^{+}-q^{+})}{q^{2}+2p_{Q}\cdot q+i\varepsilon}\delta(x_{2}p^{+}-k_{2}^{+}+q^{+}).\label{eq:VQdtotalAmp}
\end{align}
Note that the loop momentum appears inside a delta‑function in the integrand. For the same reason as before, this integral vanishes in the hard‑collinear region. In the soft‑collinear region, this integral is reduced to
\begin{align}
    \frac{\alpha_sC_F}{4\pi}V_{Qd,\rm sc}^{c}=&\frac{C_{F}}{2}(ig_{s}^{2})\delta(x_{1}-x_{1,0})\frac{x_{2}m_{Q}}{x_{3,0}}\nonumber\\
    &\times\int\frac{D^{d}q}{(2\pi)^{4}}\frac{1}{q^{2}+i\varepsilon}\frac{1}{(q-k_{2})^{2}+i\varepsilon}\frac{1}{v\cdot q+i\varepsilon}\delta\left(x_{2}-x_{2,0}+\frac{q^{+}}{p^{+}}\right).\label{eq:VQdSCETsc}
\end{align}
In the peak region, where $x_{1,0} \sim x_{2,0} \ll 1$, one can approximate $x_{3,0} \approx 1$ and replace $m_Q$ by $m_H$ to leading‑power accuracy. On the other hand, the amplitude at bHQET is
\begin{align}
\frac{\alpha_{s}C_{F}}{4\pi}V_{Qd}^{h}
=& \frac{C_{F}}{2}(ig_{s}^{2})\delta(\omega_{1}-\nu_{1})\omega_{2}\nonumber\\
    &\times\int\frac{D^{d}q}{(2\pi)^{4}}\frac{1}{q^{2}+i\varepsilon}\frac{1}{(q-k_{2})^{2}+i\varepsilon}\frac{1}{v\cdot q+i\varepsilon}\delta\left(\omega_{2}-\nu_{2}+\frac{q^{+}}{v^{+}}\right).\label{eq:VQdbHQET}
\end{align}
It is then straightforward to verify that Eq.~\eqref{eq:VQdSCETsc} and Eq.~\eqref{eq:VQdbHQET} satisfy the relation in Eq.~\eqref{eq:SCETsoftbHQET}.

Having verified the results stated in Eq.~\eqref{eq:SCEThcVanish} and Eq.~\eqref{eq:SCETsoftbHQET}, we can now apply the matching formula Eq.~\eqref{eq:jetMatchingFormula} together with the SCET and bHQET amplitudes: Eqs.~(\ref{eq:WQQhcResult}), (\ref{eq:quarkResidues}), (\ref{eq:ZSCET}) and (\ref{eq:ZbHQET}) derived in the preceding sections to extract the jet function at ${\cal O}(\alpha_s)$:
\begin{align}
    {\cal J}^{(1)}(x_{1},x_{2};\omega_{1},\omega_{2})=\delta\left(x_{1}-\frac{\omega_{1}}{m_{H}}\right)\delta\left(x_{2}-\frac{\omega_{2}}{m_{H}}\right)\left[\frac{1}{2}{\rm ln}^{2}\frac{\mu^{2}}{m_{Q}^{2}}+\frac{1}{2}{\rm ln}\frac{\mu^{2}}{m_{Q}^{2}}+\frac{\pi^{2}}{12}+2\right].
\end{align}
At leading power, the heavy quark mass $m_Q$ can be replaced by the hadron mass $m_H$. The complete expression for the jet function in the peak region, $\mathcal{J}_{\text{peak}}$, is therefore given by
\begin{align}
{\cal J}_{\rm peak}(x_{1},x_{2};\omega_{1},\omega_{2})=&\theta\left(m_{H}-\omega_{1}-\omega_{2}\right)\delta\left(x_{1}-\frac{\omega_{1}}{m_{H}}\right)\delta\left(x_{2}-\frac{\omega_{2}}{m_{H}}\right)\nonumber\\
&\times\left[1+\frac{\alpha_{s}C_{F}}{4\pi}\left(\frac{1}{2}{\rm ln}^{2}\frac{\mu^{2}}{m_{H}^{2}}+\frac{1}{2}{\rm ln}\frac{\mu^{2}}{m_{H}^{2}}+\frac{\pi^{2}}{12}+2\right)\right]. \label{eq:Jpeakto1loop}
\end{align}
The terms in square brackets are found to be identical to the corresponding ones in the jet function for the heavy meson LCDA derived in Ref.~\cite{Beneke:2023nmj}. It is not a coincidence that the jet function for $\Lambda_Q$ is identical to that for a heavy meson, and this result can be readily understood. In the one-loop correction, the heavy quark can only connect to at most one light quark—say, the $d$ quark. Although the amplitude in Eqs.~(\ref{eq:WQexpression}), (\ref{eq:Wqexpression}) and (\ref{eq:VQexpression}) also depends on the momentum fraction $x_1, x_{1,0}$ of the $u$ quark, in the peak region limit where $x_1 \approx 0$ and $x_{1,0} \approx 0$, the amplitude depends only on the momentum fractions of the heavy quark $Q$ and the $d$ quark. In this limit, the momentum fractions of $d, Q$ satisfy $x_{2}+x_3=1$ and $x_{2,0}+x_{3,0}=1$. Upon renaming $x_{2}\to u$, $x_{2,0}\to s$, and $x_{3,0}\to \bar{s}$, the resulting amplitudes become analogous to—though not identical with—the one-loop amplitudes for the heavy meson case presented in Ref.~\cite{Beneke:2023nmj}. Crucially, the differences between them vanish in the peak region limit, leading to identical jet functions after matching.

\begin{figure}
\begin{center}
\includegraphics[width=0.7\columnwidth]{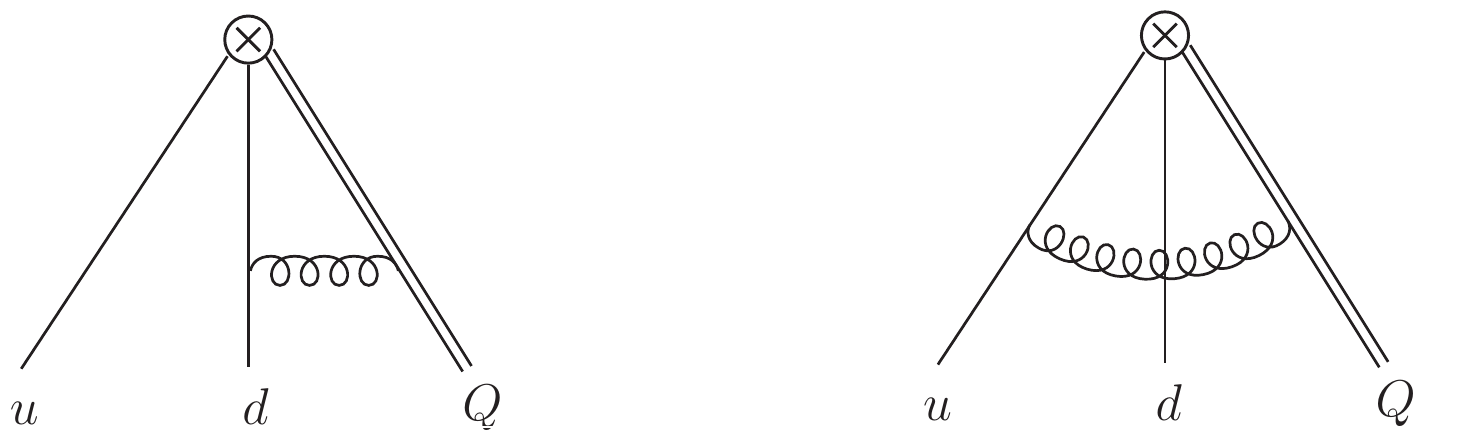} 
\caption{One-loop corrections to the matrix elements induced by local operator in Eq.~(\ref{eq:localOperMat}), where the white crossed dot denotes the local operator.}
\label{fig:LCDAmatchingLocal} 
\end{center}
\end{figure}

From the definition of the LCDAs in Eq.~(\ref{eq:defQCDLCDA}) and Eq.~(\ref{eq:defHQETLCDA}), the factorization formula relating the QCD LCDA and bHQET LCDA reads as
\begin{align}
    \Phi_{\rm QCD}(x_1,x_2)=\int d\omega_1d\omega_2\frac{{\bar f}_{\Lambda_Q}}{f_{\Lambda_Q}}{\cal J}_{\rm peak}(x_{1},x_{2};\omega_{1},\omega_{2})\Phi_{\rm bHQET}(\omega_{1},\omega_{2}).\label{eq:factorFormulaQCDHQET}
\end{align}
To derive the complete one-loop matching kernel between $\Phi_{\mathrm{SCET}}$ and $\Phi_{\mathrm{bHQET}}$, it is necessary to determine the ratio of normalization constants, ${\bar f}_{\Lambda_Q}/f_{\Lambda_Q}$, at the same order. This ratio can be obtained by matching the matrix elements in Eqs.~(\ref{eq:SCETlocalMatrix}) and (\ref{eq:bHQETlocalMatrix}) while replacing the external baryon state with free quark states
\begin{align}
f_{\Lambda_{Q}}^{\prime}&=\epsilon_{ijk}\langle0|u_i^T(0)\Gamma d_j(0)Q_{k}(0)|{\overline {u(0)d(0)}}Q(p_{Q})\rangle,\nonumber\\
{\bar f}_{\Lambda_{Q}}^{\prime}&=\epsilon_{ijk}\langle0|u_i^T(0)\Gamma d_j(0)h_{n,k}(0)|{\overline {u(0)d(0)}}Q(p_{Q})\rangle.\label{eq:localOperMat}
 \end{align}
The light quark momenta are set to zero, which does not affect the matching result.
The relevant vertex diagrams are presented in Fig.~\ref{fig:LCDAmatchingLocal}. To perform the matching, these diagrams should be calculated within both the SCET and bHQET. Note that diagrams in which the gluon connects the two light quarks are omitted, because they are identical in SCET and bHQET and therefore do not contribute to the ratio ${\bar f}_{\Lambda_Q}/f_{\Lambda_Q}$. For the same reason, contributions from the light-quark pole residues are also excluded from the calculation.

At tree level, the amplitudes coincide trivially: $f_{\Lambda_{Q}}^{\prime(0)} = {\bar f}_{\Lambda_{Q}}^{\prime(0)} = S^{(0)}$. At one-loop order, the SCET amplitude of $f_{\Lambda_{Q}}^{\prime}$ and the corresponding UV part is 
\begin{align}
f_{\Lambda_{Q}}^{\prime(1)}&=\frac{\alpha_{s}C_{F}}{4\pi}S^{(0)}\left(-\frac{1}{2\epsilon}-\frac{1}{2}{\rm ln}\frac{\mu^{2}}{m_{Q}^{2}}-1\right),\nonumber\\
f_{\Lambda_{Q},UV}^{\prime(1)}&=\frac{\alpha_{s}C_{F}}{4\pi}S^{(0)}\frac{1}{2\epsilon}.\label{eq:VlocalSCET}
\end{align}
The renormalization factor for the local operator in SCET is
\begin{align}
    Z_{c,\rm local}^{(1)}=-f_{\Lambda_{Q},UV}^{\prime(1)}-\frac{\alpha_{s}C_{F}}{4\pi}S^{(0)}\frac{1}{2}Z_{Q}^{UV}\frac{1}{\epsilon}=0.
\end{align}
Thus the renormalized $f_{\Lambda_{Q}}^{\prime(1)}$ is
\begin{align}
f_{\Lambda_{Q},\rm ren}^{\prime(1)}= & f_{\Lambda_{Q}}^{\prime(1)}+\frac{\alpha_{s}C_{F}}{4\pi}S^{(0)}\frac{1}{2}Z_{Q}^{(1)}+Z_{c,\rm local}^{(1)}= \frac{\alpha_{s}C_{F}}{4\pi}S^{(0)}\left(-\frac{2}{\epsilon}-2{\rm ln}\frac{\mu^{2}}{m_{Q}^{2}}-3\right).\label{eq:fReno}
\end{align}
After renormalization, the remaining $1/\epsilon$ singularitie are of purely IR origin. These IR divergences will be canceled by corresponding IR divergences on the bHQET side.
At one-loop order, the bHQET amplitude of ${\bar f}_{\Lambda_{Q}}^{\prime}$ vanishes due to scaleless momentum integral, while its UV part is 
\begin{align}
    {\bar f}_{\Lambda_{Q},UV}^{\prime(1)}=\frac{\alpha_{s}C_{F}}{4\pi}S^{(0)}\frac{1}{\epsilon}.
\end{align}
Thus the renormalized ${\bar f}_{\Lambda_{Q},}^{\prime(1)}$ is
\begin{align}
{\bar f}_{\Lambda_{Q},\rm ren}^{\prime(1)}=Z_{h,\rm local}^{(1)} & =-{\bar f}_{\Lambda_{Q},UV}^{\prime(1)}-\frac{\alpha_{s}C_{F}}{4\pi}S^{(0)}\frac{1}{2}Z_{h}^{UV}=-\frac{\alpha_{s}C_{F}}{4\pi}S^{(0)}\frac{2}{\epsilon},\label{eq:fbarReno}
\end{align}
where $Z_{h,\rm local}^{(1)}$ is the renormalization factor for the local operator in bHQET. Note that this result exhibits precisely the same infrared divergence as that appearing on the SCET side in Eq.~\eqref{eq:fReno}.
Using Eq.~(\ref{eq:fReno}) and Eq.~(\ref{eq:fbarReno}), one obtains the ratio of normalization constants as:
\begin{align}
    \frac{{\bar f}_{\Lambda_Q}}{f_{\Lambda_Q}}=\frac{{\bar f}_{\Lambda_{Q}}^{\prime(0)}+{\bar f}_{\Lambda_{Q},\rm ren}^{\prime(1)}}{f_{\Lambda_{Q}}^{\prime(0)}+f_{\Lambda_{Q},\rm ren}^{\prime(1)}}=1+\frac{\alpha_{s}C_{F}}{4\pi}\left(2{\rm ln}\frac{\mu^{2}}{m_{Q}^{2}}+3\right).
\end{align}
Inserting this result into Eq.~\eqref{eq:factorFormulaQCDHQET} and taking $m_Q \approx m_H$ yields the matching relation between $\Phi_{\mathrm{QCD}}$ and $\Phi_{\mathrm{bHQET}}$:
\begin{align}
    &\Phi_{\mathrm{QCD}}(x_1,x_2)\nonumber\\
    =&\ m_H^2\left[1+\frac{\alpha_{s}C_{F}}{4\pi}\left(\frac{1}{2}{\rm ln}^{2}\frac{\mu^{2}}{m_{H}^{2}}+\frac{5}{2}{\rm ln}\frac{\mu^{2}}{m_{H}^{2}}+\frac{\pi^{2}}{12}+5\right)\right]\Phi_{\mathrm{bHQET}}(x_1 m_H, x_2 m_H). \label{eq:FacFormulaLCDA}
\end{align}
In the two-step matching scheme, once the QCD LCDA is extracted from the quasi-DA, one can then use this formula to determine the corresponding HQET LCDA.

\section{Conclusion}\label{sec:conclusion}

In this work, we have established a factorization framework connecting the leading-twist QCD LCDA of the $\Lambda_Q$ baryon to its counterpart in bHQET in the peak region. This relation serves as a crucial bridge for extracting the non-perturbative HQET LCDA from first-principles lattice QCD calculations, thereby enabling precision phenomenology in heavy baryon decays.
We computed the one-loop perturbative corrections to both the QCD and bHQET LCDAs, and demonstrated that the matching procedure can be substantially simplified by employing the method of regions. From this matching, we derived the jet function that acts as the perturbative matching kernel in the factorization formula. The jet function obtained for the \(\Lambda_Q\) baryon exhibits precisely the same logarithmic structure as the jet function for the heavy meson LCDA. Our result constitutes a first essential step toward a non-perturbative determination of the \(\Lambda_Q\) LCDA from lattice QCD, and provides a solid foundation for further theoretical developments and phenomenological applications in heavy baryon weak decays.

\section*{Acknowledgements}
We thank Yan-Bin Wei and Wei Wang for valuable discussions. This work is supported in part by Natural Science Foundation of China under Grants No. 12565014, 12305103, 12205180. The work of Jun Zeng is also supported by the Talent Research Startup Foundation of Hainan Normal University: HSZK-KYQD-202523, and Opening Foundation of Shanghai Key Laboratory of Particle Physics and Cosmology under Grant No. 22DZ2229013-5.

\begin{appendix}
	
\section{Complete amplitudes in SCET}\label{sec:CompleteAmpSCET}
This appendix presents the full SCET amplitudes of all the diagrams in Fig.~\ref{fig:LCDAmatching}, with the exception of $W_{QQ}^c$, which has already been given in Eq.~\eqref{eq:WQQhcResult}.
\begin{align}
W_{Qd}^c=&\frac{1}{2}\frac{\alpha_{s}C_{F}}{4\pi}\delta(x_{1}-x_{1,0})\theta(\bar{x}_{1,0}-x_{2})\theta(x_{2}-x_{2,0})S^{(0)}\nonumber\\
&\times\left(\frac{\mu^{2}}{(4\pi)e^{-\gamma_{E}}}\right)^{\epsilon}\frac{(16\pi^{2})\pi^{1-\epsilon}}{(2\pi)^{D-1}}m_{Q}^{-2\epsilon}\Gamma(\epsilon)\frac{\bar{x}_{1,0}-x_{2}}{(x_{2}-x_{2,0})^{1+2\epsilon}x_{3,0}^{1-2\epsilon}},\\
W_{Qu}^c=&\frac{1}{2}\frac{\alpha_{s}C_{F}}{4\pi}\delta(x_{2}-x_{2,0})\theta(\bar{x}_{2,0}-x_{1})\theta(x_{1}-x_{1,0})S^{(0)}\nonumber\\
&\times\left(\frac{\mu^{2}}{(4\pi)e^{-\gamma_{E}}}\right)^{\epsilon}\frac{(16\pi^{2})\pi^{1-\epsilon}}{(2\pi)^{D-1}}m_{Q}^{-2\epsilon}\Gamma(\epsilon)\frac{\bar{x}_{2,0}-x_{1}}{(x_{1}-x_{1,0})^{1+2\epsilon}x_{3,0}^{1-2\epsilon}},\\
V_{Qd}^c=& \frac{1}{2}\frac{\alpha_{s}C_{F}}{4\pi}\delta(x_{1}-x_{1,0})\theta(x_{2,0}-x_{2})S^{(0)}\nonumber\\
&\times\left(\frac{\mu^{2}}{(4\pi)e^{-\gamma_{E}}}\right)^{\epsilon}\frac{(16\pi^{2})\pi^{1-\epsilon}\Gamma(\epsilon)}{(2\pi)^{D-1}}m_{Q}^{-2\epsilon}\frac{x_{2}}{2x_{2,0}}\left[\frac{1}{1-x_{1}}+\frac{1}{x_{2,0}-x_{2}}\right]\left(\frac{-x_{2}(x_{2,0}-x_{2})}{x_{3,0}(1-x_{1})}\right)^{-\epsilon}\nonumber\\
+ & \frac{1}{2}\frac{\alpha_{s}C_{F}}{4\pi}\delta(x_{1}-x_{1,0})\theta(x_{2,0}+x_{3,0}-x_{2})\theta(x_{2}-x_{2,0})S^{(0)}\left(\frac{\mu^{2}}{(4\pi)e^{-\gamma_{E}}}\right)^{\epsilon}\nonumber\\
& \times\frac{(16\pi^{2})\pi^{1-\epsilon}\Gamma(\epsilon)}{(2\pi)^{D-1}}m_{Q}^{-2\epsilon}\frac{1}{2}\left\{ \frac{x_{2}}{x_{2,0}}\frac{2}{x_{2}-x_{2,0}}\left[\left(\frac{\bar{x}_{1,0}-x_{2,0}}{x_{2}-x_{2,0}}\right)^{2\epsilon}-\left(\frac{(\bar{x}_{1,0}-x_{2,0})\bar{x}_{1,0}}{x_{2}(x_{2}-x_{2,0})}\right)^{\epsilon}\right]\right.\nonumber\\
 &\left.+\left[\frac{1}{\bar{x}_{1,0}}\frac{x_{2}}{x_{2,0}}\left(\frac{(\bar{x}_{1,0}-x_{2,0})\bar{x}_{1,0}}{x_{2}(x_{2}-x_{2,0})}\right)^{\epsilon}-\frac{x_{2}-x_{2,0}}{x_{2,0}(\bar{x}_{1,0}-x_{2,0})}\left(\frac{\bar{x}_{1,0}-x_{2,0}}{x_{2}-x_{2,0}}\right)^{2\epsilon}\right]\right\},\\
 V_{Qu}^c=& \frac{1}{2}\frac{\alpha_{s}C_{F}}{4\pi}\delta(x_{2}-x_{2,0})\theta(x_{1,0}-x_{1})S^{(0)}\nonumber\\
 &\times\left(\frac{\mu^{2}}{(4\pi)e^{-\gamma_{E}}}\right)^{\epsilon}\frac{(16\pi^{2})\pi^{1-\epsilon}\Gamma(\epsilon)}{(2\pi)^{D-1}}m_{Q}^{-2\epsilon}\frac{x_{1}}{2x_{1,0}}\left[\frac{1}{1-x_{2}}+\frac{1}{x_{1,0}-x_{1}}\right]\left(\frac{-x_{1}(x_{1,0}-x_{1})}{x_{3,0}(1-x_{2})}\right)^{-\epsilon}\nonumber\\
+ & \frac{1}{2}\frac{\alpha_{s}C_{F}}{4\pi}\delta(x_{2}-x_{2,0})\theta(x_{1,0}+x_{3,0}-x_{1})\theta(x_{1}-x_{1,0})S^{(0)}\left(\frac{\mu^{2}}{(4\pi)e^{-\gamma_{E}}}\right)^{\epsilon}\nonumber\\
&\times\frac{(16\pi^{2})\pi^{1-\epsilon}\Gamma(\epsilon)}{(2\pi)^{D-1}}m_{Q}^{-2\epsilon}\frac{1}{2}\left[\frac{x_{1}}{x_{1,0}}\left(\frac{1}{x_{1,0}+x_{3,0}}-\frac{2}{x_{1}-x_{1,0}}\right)\left(\frac{x_{1}(x_{1}-x_{1,0})}{x_{3,0}(x_{1,0}+x_{3,0})}\right)^{-\epsilon}\right.\nonumber\\
 &\left.+\left(\frac{2x_{1}}{x_{1,0}(x_{1}-x_{1,0})}-\frac{x_{1}-x_{1,0}}{x_{3,0}}\right)\left(\frac{x_{1}-x_{1,0}}{x_{3,0}}\right)^{-2\epsilon}\right],\\
 W_{uu}^c=&W_{dd}^c=W_{ud}^c=W_{du}^c=W_{uQ}^c=W_{dQ}^c=V_{ud}^c=0.
\end{align}

\section{Complete amplitudes in bHQET}\label{sec:CompleteAmpbHQET}
This appendix presents the full bHQET amplitudes of all the diagrams in Fig.~\ref{fig:LCDAmatching}.  
\begin{align}
W_{QQ}^c=&0,\nonumber\\
W_{Qd}^c=&\frac{1}{2}\frac{\alpha_{s}C_{F}}{4\pi}\left(\frac{\mu^{2}}{(4\pi)e^{-\gamma_{E}}}\right)^{\epsilon}\frac{(16\pi^{2})\pi^{1-\epsilon}\Gamma(\epsilon)}{(2\pi)^{D-1}}S^{(0)}\delta(\omega_{1}-\nu_{1})\frac{\theta(\omega_{2}-\nu_{2})}{(\omega_{2}-\nu_{2})^{1+2\epsilon}},\\
W_{Qu}^c=&\frac{1}{2}\frac{\alpha_{s}C_{F}}{4\pi}\left(\frac{\mu^{2}}{(4\pi)e^{-\gamma_{E}}}\right)^{\epsilon}\frac{(16\pi^{2})\pi^{1-\epsilon}\Gamma(\epsilon)}{(2\pi)^{D-1}}S^{(0)}\delta(\omega_{2}-\nu_{2})\frac{\theta(\omega_{1}-\nu_{1})}{(\omega_{1}-\nu_{1})^{1+2\epsilon}},\\
V_{Qd}^c=&\frac{1}{2}\frac{\alpha_{s}C_{F}}{4\pi}\delta(\omega_{1}-\nu_{1})S^{(0)}\left(\frac{\mu^{2}}{(4\pi)e^{-\gamma_{E}}}\right)^{\epsilon}\frac{(16\pi^{2})\pi^{1-\epsilon}\Gamma(\epsilon)}{(2\pi)^{D-1}}\frac{\omega_{2}}{\nu_{2}}\nonumber\\
&\times\left\{ e^{i\pi\epsilon}\omega_{2}^{-\epsilon}\frac{\theta(\nu_{2}-\omega_{2})}{(\nu_{2}-\omega_{2})^{1+\epsilon}}+\frac{\theta(\omega_{2}-\nu_{2})}{(\omega_{2}-\nu_{2})^{1+\epsilon}}\left[(\omega_{2}-\nu_{2})^{-\epsilon}-\omega_{2}^{-\epsilon}\right]\right\},\\
V_{Qu}^c=&\frac{1}{2}\frac{\alpha_{s}C_{F}}{4\pi}\delta(\omega_{2}-\nu_{2})S^{(0)}\left(\frac{\mu^{2}}{(4\pi)e^{-\gamma_{E}}}\right)^{\epsilon}\frac{(16\pi^{2})\pi^{1-\epsilon}\Gamma(\epsilon)}{(2\pi)^{D-1}}\frac{\omega_{1}}{\nu_{1}}\nonumber\\
&\times\left\{ e^{i\pi\epsilon}\omega_{1}^{-\epsilon}\frac{\theta(\nu_{1}-\omega_{1})}{(\nu_{1}-\omega_{1})^{1+\epsilon}}+\frac{\theta(\omega_{1}-\nu_{1})}{(\omega_{1}-\nu_{1})^{1+\epsilon}}\left[(\omega_{1}-\nu_{1})^{-\epsilon}-\omega_{1}^{-\epsilon}\right]\right\},\\
W_{uu}^h=&W_{dd}^h=W_{ud}^h=W_{du}^h=W_{uQ}^h=W_{dQ}^h=V_{ud}^h=0.
\end{align}

\end{appendix}

\end{document}